# USE OF TWO-PARTICLE SIMULATIONS FOR THE EVALUATION OF THE FLOCCULATION RATE IN THE CASE OF INSURMOUNTABLE REPULSIVE BARRIERS

German Urbina-Villalba

Instituto Venezolano de Investigaciones Científicas (IVIC), Centro de Estudios Interdisciplinarios de la Física (CEIF),
Carretera Panamericana Km. 11, Aptdo. 20632, Caracas, Venezuela. Email: guv@ivic.gob.ve

**Abstract**    The interaction potential between two colloidal particles typically spans a few nanometers. Hence the correct appraisal of the potential in simulations requires very short time steps. However, it is often possible to combine *one* many-particle calculation with several two-particle simulations in order to estimate a set of aggregation rates. As in the theory of Fuchs, this procedure combines the evaluation of the fastest flocculation rate with the calculation of several stability ratios. This methodology was already tested in systems which exhibit low repulsive barriers for primary minimum flocculation. In this article, the procedure is modified in order to calculate an average rate for secondary minimum flocculation. The technique is then used to estimate the influence of drop deformation on the aggregation rate of hexadecane-in-water nano-emulsions stabilized with sodium dodecylsulfate (SDS).

## 1. INTRODUCTION

According to the theory of Derjaguin, Landau, Verwey and Overbeek (DLVO theory) [Derjaguin, 1941; Verwey, 1948], the stability of particles suspended in a liquid essentially depends on their interaction forces. In the absence of repulsive forces, solid particles diffuse freely until they make contact. The minimum distance of approach corresponds to the lowest potential energy of the two-particle system. This process is referred to as primary minimum flocculation, and is regarded to be "irreversible" due to the strength of the van der Waals attraction at very close separations. Repulsive electrostatic forces create a potential energy barrier that hinders the attainment of primary minimum flocculation. The barrier must be surpassed in order to reach the minimum. Hence, irreversible flocculation is effectively delayed by a repulsive barrier providing kinetic stability to the dispersion.

The quantitative implementation these ideas is based on a modification of the theory of irreversible coagulation [Smoluchowski, 1917] which assumes that the particles stick irreversibly to each other as soon as they collide. This process generates clusters of particles that progressively grow as a function of time. The number concentration of aggregates of $k$ primary particles (size $k$) existing at a given time ($n_k(t)$) results from a balance between the aggregates produced by the collisions of smaller clusters of sizes $i$ and $j$ (such that $i+j=k$), and the aggregates of size $k$ lost by the collisions with clusters of any other size:

$$\frac{dn_k(t)}{dt} = \frac{1}{2}\sum_{\substack{i=1\\j=k-i}}^{i=k-1} k_{ij}\, n_i(t)\, n_j(t) - n_k(t)\sum_{i=1}^{\infty} k_{ik}\, n_i(t) \quad (1)$$

Equation (1) can only be solved for certain sets of aggregation rates ($k_{ij}$). In the simplest scenario, referred as "constant kernel" approximation: $k_{ij}=k_f$, and:

$$n_k = \frac{n_0\,(k_f\,n_0\,t)^{k-1}}{(1+k_f\,n_0\,t)^{k+1}} \quad (2)$$

Using Eq. (2) an expression for the time dependence of the total number of aggregates ($n$) per unit volume results:

$$n = \sum n_k = \frac{n_0}{1+k_f\,n_0\,t} = \frac{n_0}{1+t/\tau_c} \quad (3)$$

Here $n_0$ stands for the initial number of aggregates, and $\tau_c$ is a characteristic aggregation time equal to the half lifetime of the dispersion:

$$n\,(t=t_{1/2}) = \frac{n_0}{1+k_f\,n_0\,t_{1/2}} = \frac{n_0}{2} \quad (4)$$

$$t_{1/2} = \frac{1}{k_f\, n_0} = \tau_c \qquad (5)$$

Hence, $\tau_c$ is a direct measurement of the stability of a suspension towards flocculation. It solely depends on the initial concentration of aggregates and the flocculation rate ($k_f$).

In order to evaluate $k_f$ Smoluchowski used Fick's laws to calculate the flux of particles ($J$) that diffuses towards a central particle fixed in space:

$$J = -\frac{\partial n}{\partial t} = 4\pi r^2 \left( D\, \frac{\partial u}{\partial r} \right) \qquad (6)$$

Here u(r) is the spatial distribution of particles, and $D$ their diffusion coefficient. The friction ($\mu$) enforced by the surrounding liquid on a particle of radius $R$, is related to $D$ through the Stokes-Einstein's relation ($\mu D = k_B T$), where $k_B$, $T$ and $\eta$ stand for Boltzmann's constant, the absolute temperature of the system, and the viscosity of the continuous phase. The drag coefficient is equal to: $\mu(r) = \mu_i \beta(r/R)$, where $\mu_i$ is the value of the coefficient of Stokes at infinite dilution ($6\pi\eta R$), and $\beta$ is a dimensionless function which accounts for the hydrodynamic interaction between the particles [Derjaguin, 1967; Sonntag, 1987].

In the absence of hydrodynamic interactions ($\beta = 1$), and assuming that the particles only move as a result of their thermal interaction with the surrounding medium, a simple expression for the flocculation rate is obtained [Smoluchowski, 1917]:

$$k_f = k_s = 8\pi D\tilde{R} = \frac{4\,k_B T}{3\eta} \qquad (7)$$

In Eq. (7) $\tilde{R}$ represents the collision radius between two particles: $\tilde{R} \sim (R_i + R_j)/2$. This distance depends on the topology of the aggregates and on the probability of a collision [Lattuada, 2003]. At room temperature $k_f$ is then approximately equal to: $6 \times 10^{-18}\, m^3/s$.

Interaction forces are formally incorporated in the theory of Smoluchowski [Derjaguin, 1967; Sonntag, 1987; Fuchs, 1936] adding an additional term to Eq. (6) which takes into account the force produced by the interaction potential:

$$J = 4\pi r^2 \left( D\, \frac{\partial u}{\partial r} + \frac{u}{\mu}\, \frac{dV}{dr} \right) \qquad (8)$$

Here $V(r)$ is the free energy of interaction between two particles ($V_T = V_A + V_R$), equal to the sum of the attractive interactions between their molecules ($V_A$), and the repulsive interactions ($V_R$) between their surface charges. The solution

of this equation leads to a simple modification of the aggregation rate [Fuchs, 1936]:

$$k_f = k_f^{slow} = \frac{k_f^{fast}}{W} \qquad (9)$$

Where superscripts "fast" and "slow" refer to rapid and hindered aggregation (low and high salinity respectively, in the case of electrostatically stabilized dispersions), and $W$ is known as the *stability ratio* [Fuchs, 1936; McGown, 1967]:

$$W = \frac{\displaystyle\int_{R_i+R_j}^{\infty} [\mu(r)/r^2]\exp(V_T/k_B T)\,dr}{\displaystyle\int_{R_i+R_j}^{\infty} [\mu(r)/r^2]\exp(V_A/k_B T)\,dr} \qquad (10)$$

Thus, the half lifetime of a suspension can be calculated using Eqs. (5), (9) and (10). Repulsive potentials generate large values of $W$, and as a consequence, small aggregation rates. In the case of an electrostatic potential, the repulsive barrier decreases with the augment of the ionic strength of the solution, favoring a stability ratio closer to $W = 1$.

According to Eqs. (5) and (9):

$$W = \frac{k_f^{fast}}{k_f^{slow}} = \frac{\tau_c^{slow}}{\tau_c^{fast}} \qquad (11)$$

Eq. (11) is an approximation based on the *ansatz* that the only effect of the repulsive barrier is to delay primary minimum flocculation. According to the simulations the actual situation is more complex since equation (3) does not seem to hold for emulsions in the presence of a substantial repulsive barrier [Urbina-Villalba, 2006]. Moreover, $k_f$ and $W$ are usually evaluated experimentally from the rate of doublet formation $k_{11}$ [Sonntag, 1987; Lips, 1971; Lichtenbelt, 1973; Gregory, 2009; Young, 1991; Mendoza, 2013; Verbich, 1997]. In this case it is supposed that $k_f \sim k_{11}/2$ so that:

$$W \approx \frac{k_{11}^{fast}}{k_{11}^{slow}} \qquad (12)$$

However, the relationship between $k_f$ and $k_{11}$ ($k_f \sim k_{11}/2$) is not necessarily correct, since these constants are deduced from Eq. (1) for two limiting cases: a) assuming that aggregates of all sizes occur, or b) supposing that the process of aggregation stops with the formation of doublets. As a result an inconsistency arises: if the stoichiometry of Eq. (1) is correct and the constant kernel approximation assumes that all the members of the kernel $\{k_{ij}\}$ are equal to $k_f$. How is then possible that $k_f = k_{11}$ and at the same time $k_f = k_{11} / 2$?

If equation (12) is approximately correct, Eqs. (3, 5, 7, 9 -12) allow establishing a quantitative description of the overall stability of a suspension due to aggregation.

## 2. EXPERIMENTAL EVALUATION OF THE AGGREGATION RATE

The theory described in Section 1 strictly applies to solid particles which cannot coalesce [Urbina-Villalba, 2015]. Moreover, the experimental longtime behavior of dispersions is rarely contrasted with the predictions of the theory. Recently our group developed a new technology for the evaluation of the mixed flocculation-coalescence rate ($k_{FC}$) in emulsions. In the case of non-deformable drops the measured aggregation rate could include the process of coalescence because drops necessarily coalesce when they reach the primary minimum of the interaction potential [Danov, 1994; Urbina-Villalba, 2006]. This cannot be assured in the case of deformable droplets since the deformation of the drops implies additional repulsive barriers that hinder the process. In any event, an evaluation of the aggregation rate partially considers the finite time required for the drainage of the intermediate fluid between two droplets. Only in the case in which the film is stable for a period of time substantially larger than $\tau_c$, it is possible to separate both processes and evaluate their rates independently.

The procedure referred above relies on the formulation of a suitable theoretical expression for the turbidity ($\tau$) of a nano-emulsion as a function of time [Rahn-Chique, 2012].

$$\tau = n_1 \sigma_1 + x_d \sum_{k=2}^{k_{max}} n_k \sigma_{k,d} + (1 - x_d) \sum_{k=2}^{k_{max}} n_k \sigma_{k,s} \quad (13)$$

Here: $\sigma_1$, $\sigma_{k,d}$ and $\sigma_{k,s}$ represent the optical cross sections of primary drops, aggregates of size $k$ and spherical drops of size $k$ ($R_k = \sqrt[3]{k} \, R_0$). Fitting of Eq. (13) to the experimental data, allows the calculation of $k_{FC}$, and $x_d$. In the case of ionic nano-emulsions, aggregation is induced by injecting a high concentration of salt into the sample vessel in order to screen the surface charge of the drops produced by the adsorption of surfactant molecules.

Using equation (13) the behavior of a hexadecane in water nanoemulsion (R = 184 nm) stabilized with 0.5 mM of sodium dodecylsulfate between 300 mM and 1 M NaCl [Urbina-Villalba, 2015] was studied. Many-particle emulsion stability simulations (ESS) of these systems were also run assuming non-deformable droplets. The surface charge of the drops was parameterized either by extrapolating the variation of their zeta potential to larger salinities (it could only be measured between 1 mM < [NaCl] < 200 mM) or by using one zeta potential measurement and a set of adsorption isotherms [Urbina-Villalba, 2013]. The repulsive barriers resulting from the latter procedure (Type I parameterization (T-I)) are low favoring coalescence and fast rates. Instead, the ones resulting from zeta potential measurements (Type II parametrization, T-II) are high enough to preserve the number of particles during 40 s. Figure 1 shows the comparison between theory (T-II) and experiment. The agreement is satisfactory considering the restrictions of the simulations. Among other limitations, only *one simulation* of *500 drops* was used for each ionic strength. Equilibrium surface excesses were assumed in all cases, disregarding the effect of dynamic adsorption and surfactant redistribution in the stability of the emulsions. Additionally, concentration-independent (average) coefficients were used in the theoretical expression of the surface excess as a function of the salt concentration (Eq. (29) of Ref. [Urbina-Villalba, 2013]).

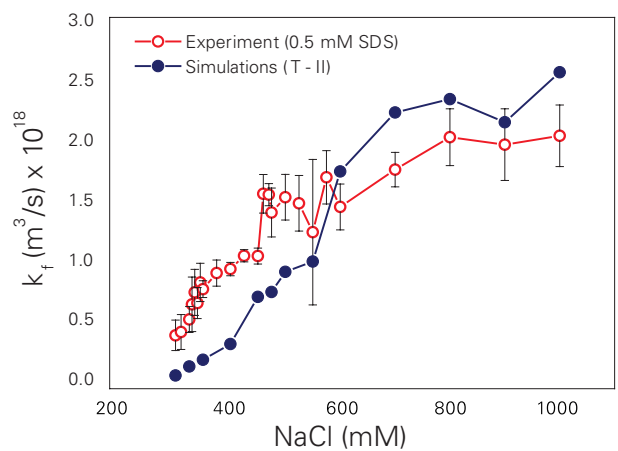

Fig. 1: Salt dependence of the flocculation rate of hexadecane-in-water nano-emulsions for a surfactant concentration of 0.5 SDS. Theoretical results for non-deformable spherical drops correspond to 500-particle simulations (T - II).

Since only secondary minimum flocculation was observed, it is not possible to reconcile the quantitative predictions of the DLVO theory with the calculations. The simulations indicate that for the referred system, the repulsive barrier is not crossed even at 1 M NaCl. This is a remarkable finding since it suggests that the theoretical formalism used by DLVO to make quantitative predictions cannot be applied (at least in this case).

According ESS the variation of $k_{FC}$ with the ionic strength results from the reversible aggregation of the drops within flocs of any size. Reversible aggregation causes the average size of each cluster to fluctuate in time due to its partial disintegration. This fluctuation of the cluster size induces a decrease in the temporal variation of the total number of aggregates, and as a result, a smaller flocculation rate. As the ionic strength increases, the depth of the secondary minimum between two drops increases, and the position of the minimum shifts towards shorter separation distances. These two factors increase the stability of the aggregates favoring faster aggregation rates. The increase approaches a limit value when the position of the secondary minimum gets closer to the minimum separation distance between two particles in the secondary minimum.

## 3. ESCAPE OVER A POTENTIAL BARRIER

It is clear that the quantitative predictions of DLVO rely on the ability of the particles to reach the primary minimum. Equations (9) and (10) assume that this minimum will always be attained despite the magnitude of the repulsive barrier. The assumptions implicit in the formulation of the stochastic equations that justify this behavior are somewhat hidden within the mathematical framework that describes the evolution of the system in the phase space [Mazo, 2002]. Most treatments are classical though this is a microscopic system and the "Tunnel phenomenon" is a well established fact in Quantum Mechanics.

Kramers [Kramers, 1940] specifically studied the problem of Brownian motion in a field of force. He deduced explicit expressions for the process of "jumping over" a potential barrier of size Q separating two states of energy A and B. Whenever the effect of the "Brownian forces" on the velocity of the particle is much larger than that of the *external* force, the reaction velocity for the process is expected to be:

$$k_r = \frac{2 \pi \omega \omega'}{\eta} e^{-Q/k_B T} \tag{14}$$

Where $\omega$ (and $\omega'$) correspond to the amplitude of the harmonic potentials used to represent the potential energy of state A:

$$U = \frac{1}{2} (2 \pi \omega)^2 q^2 \tag{15}$$

and the maximum of the potential barrier (Q) located at $q_c$ along the "reaction" coordinate ($q$):

$$U = Q - \frac{1}{2} (2 \pi \omega')^2 (q - q_c)^2 \tag{16}$$

A different expression results if the viscosity of the solutions is substantially smaller than the external force (when the Brownian forces do not change the velocity of the particle appreciably):

$$k_r = \eta \frac{Q}{k_B T} e^{-Q/k_B T} \tag{17}$$

As shown by Kramers, Eqs. (14) and (17) are consistent with the Transition State Theory enunciated by Eyring [Eyring, 1935]. In both cases the effect of the barrier is to delay the passing from A to B as envisaged by Fuchs (Eqs. (9) and (10)). In fact, the numeric evaluation of Eq. (10), led Prieve and Ruckenstein [Prieve, 1977] to deduce a direct relationship between the height of the repulsive barrier and the stability ratio. Such functionality was later reproduced for emulsions using ESS [Lozsán, 2006]:

$$\ln W = 0.493 \,(\Delta V / k_B T - 1) \tag{18}$$

Where: $\Delta V$ (~ Q) is the difference in energy between the maximum of the repulsive barrier and the bottom of the secondary minimum.

According to Langevin [Langevin, 1908], a particle suspended in a fluid is expected to exhibit ballistic motion at short times, with an instantaneous thermal velocity of:

$$\frac{\langle x^2 \rangle^{1/2}}{t} = \sqrt{k_B T/m} \tag{19}$$

Where m is the mass of the particle. In this regime, the collisions of the solvent molecules with the particle are insufficient to change its velocity appreciably. On the other hand, when $t \gg m / 6\pi\eta R = \tau_p$ (where R is the radius of the particle) the number of collisions is huge, and the displacement of the particle changes erratically in magnitude and position. Hence, the velocity is not well defined due to the fractal trajectory of the particle [Einstein, 1948]:

$$\frac{\langle x^2 \rangle^{1/2}}{t} = \frac{\sqrt{2Dt}}{t} \propto t^{-1/2} \tag{20}$$

Recent measurements of the instantaneous velocity employing optical tweezers showed that the ballistic regime can be accurately reproduced. Thermal velocities of 0.422 - 0.425 mm/s are exhibited by a $SiO_2$ particle of 2.8 μm in air (expected: 0.429 mm/s) [Li, 2010], and 0.35 mm/s in water (expected: 0.36 mm/s) [Kheifets, 2014]. In the case of air, values of $\tau_p$ equal to 48.5 (99.8 kPa) and 147.3 μs (2.75 kPa) were deduced from the fitting of the normalized velocity autocorrelation function of the particles in the optical trap. Slight deviations of the mean square displacement from the ballistic regime were observed above $10^{-5}$ s, but the regime appears to extend from $10^{-6}$ to (at least) $10^{-4}$ s [Li, 2010]. In the case of water, $\tau_p = 1.2$ μs. Thus, the ballistic regime was confirmed from $10^{-8}$ s to a fraction of $\tau_p$ [Kheifets, 2014]. Hydrodynamic memory effects generate an intermediate period between purely ballistic and solely diffusive, where the mean square displacement (MSD) takes a rather complicated form. The correct hydrodynamic treatment requires the use of an effective mass in the calculation of the MSD in Eq. (19) equal to the sum of the mass of the particle and half the mass of the displaced fluid [Huang, 2011]. Using these corrections a thermal velocity of 11 mm/s is predicted for a 184-nm drop of hexadecane in water, along with a relaxation time of $\tau_p = 6.6 \times 10^{-9}$ s. This corresponds to a kinetic energy of only $0.3 \, k_B T$.

Since the average energy of a thermal bath is $k_B T$, the energy required for the particles for crossing a repulsive barrier of higher height can only be provided by fluctuations of its kinetic energy induced by the mechanism of momentum exchange. The maximum number of collisions between a drop and the solvent molecules per unit time can be estimated as the ratio between the area of the drop and the cross section of a water molecule [Urbina-Villalba, 2003]. Using the equipartition theorem, the maximum momentum transferred by one water molecule can be estimated. The total momentum transferred results from the unbalance between the collisions received in every direction. For a particle of 1 μm, an unbalance of 1% generates a maximum energy transfer of 15 $k_B T$ [Urbina-Villalba, 2003]. The same percentage can only induce an energy transfer of 3 $k_B T$ and a velocity of 34.7 mm/s on a drop of hexadecane of 184 nm.

## 4. USE OF THE CHARACTERISTIC TIME OF DOUBLET FORMATION FOR THE APPRAISAL OF SECONDARY-MINIMUM FLOCCULATION

Unfortunately it is always possible to justify a given experimental value of the aggregation rate in terms of a suitable stability ratio. However, in the case of emulsions, primary minimum flocculation may result in the coalescence of the drops. Hence, coalescence –whenever detectable- serves as a validation of surmounting the repulsive barrier.

In the case of non-deformable drops, the code of emulsion stability simulations automatically coalesces the particles when primary minimum flocculation is reached. Thus, it is possible to calculate stability ratios using coagulation (coalescence) times ($\tau_{coal}$) even in the case of solid particles. This modus operandi reduces the number of many-particle simulations to the minimum:

$$k_{FC}^{slow} = \frac{k_{FC}^{fast}}{W} = \left( \frac{\tau_c^{fast}}{\tau_c^{slow}} \right) k_{FC}^{fast} \approx \left( \frac{\tau_{coal}^{fast}}{\tau_{coal}^{slow}} \right) k_{FC}^{fast}$$

(21)

The procedure outlined above was implemented in Ref. [Urbina-Villalba, 2009a] in order to evaluate the aggregation rate of 96-nm anionic latex particles suspended in aqueous solution (450 < [NaCl] < 750 mM). The width of the repulsive barrier of these systems was lower than 3 nm. Hence, a time step of $4.5 \times 10^{-11}$ s was used. The fast aggregation rate was computed using only 125 drops. The coalescence times were obtained using small cubic cells with L = 5 R. Each pair of particles was initially separated by only 20 nm. The calculation ended as soon as the particles coalesced. Each simulation was repeated 100 times in order to obtain a representative average value with the lowest computational effort. The simulations differed from each other in the initial seed of the random number generator, which was changed in order to induce distinct aggregation paths. The value of $\tau_{coal}$ for fast aggregation was evaluated using only the van der Waals potential between the particles. The rest of the computations included the corresponding repulsive interaction. The agreement between theory and experiment was good except at low ionic strength where the slope of logW vs. [NaCl] increased much stronger than experimentally found, indicating than the potential barrier was "harder" than expected. However, this slope was approximately reproduced by many-particle simulations using the same potential (see Fig. 3c in [Urbina-Villalba, 2009a]). It was also observed that repulsive barriers larger than 10 $k_B T$ could not be surmounted by particles of this size [Lozsán, 2006, Urbina-Villalba, 2009a, Urbina-Villalba, 2009b]. As a result, the procedure was limited to salinities above 400 mM.

Equation (21) was also employed to study the effect of drop deformability on the stability of decane-in-water emulsions stabilized with either AOT or sodium oleate. In this case, the agreement between the simulations and the experimental data was fairly good for all ionic strengths [Osorio, 2011]. However, it was concluded that the experimental data was insufficient in order to discriminate between deformable and non-deformable drops.

Nowadays, there is a significant number of articles devoted to understand the role of secondary minimum flocculation on the stability of emulsions. This minimum generally occurs at a distance of approach substantially larger than the position of the repulsive barrier. As a result, the minimum is usually shallow, meaning that the particles can go in and out without a considerable cost of energy. Hence, secondary minimum aggregation is often regarded as a reversible phenomenon. It decreases the population of doublets required for the building of larger aggregates, and therefore it effectively lowers the value of $k_f$. Recent theoretical studies even suggest that secondary minimum flocculation may also lead to the observance of two characteristic times for doublet formation [Ohshima, 2014a; Ohshima, 2014b].

In the past, our group studied the possible occurrence of *fast* secondary minimum flocculation [Urbina-Villalba, 2005]. Such phenomenon is likely to happen between particles of micron size at high ionic strength, where the secondary minimum is deep. In this case, the potential energy curve closely follows the van der Waals potential until the minimum is reached. Depending on the position of the minimum, its depth, and the particle size, fluctuations in the total number of aggregates of every size are observed. These are caused by the reversible aggregation of the particles. The stability of the aggregates increases with the depth of the minimum, which grows deeper as the ionic strength increases.

In this article we establish an ad-hoc criterion for secondary minimum flocculation based on the time of residence of the particles within the secondary minimum. This convention allows the calculation of stability ratios for those cases in which primary minimum flocculation is not possible (the potential barriers between spherical drops vary between 60 and 506 $k_B T$).

Using the procedure outlined in the previous paragraph, the qualitative behavior of the flocculation rate for 0.5 SDS was successfully reproduced using non-deformable droplets. Once the procedure has been validated, an attempt to explain the variation of the flocculation rates corresponding to a surfactant concentration of 7.5 mM SDS was made. In this case, many-particle simulations of non-deformable drops completely fail (Figure 2). The experimental flocculation rate appears to increase up to 500 mM reaching a short plateau, and then slightly decreases from 600 mM up to 1 M.

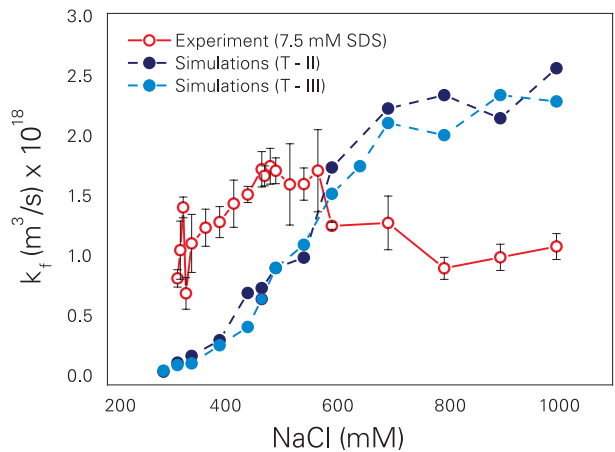

**Fig. 2:** Salt dependence of the flocculation rate of hexadecane-in-water nanoemulsions for a surfactant concentration of 7.5 mM. Theoretical results correspond to 500-particle simulations of spherical drops. Similar results are obtained using either average (T - II) or concentration-dependent coefficients (T - III) in the calculation of the surface excess (Eq. (29) of Ref. [Urbina-Villalba, 2015]).

Since high surfactant concentrations favor lower interfacial tensions, here we consider the possible influence of drop deformation on the aggregation rate using two-particle simulations. Moreover, the possible occurrence of alternative coalescence mechanisms other than film drainage (like capillary waves and hole formation) is also studied.

## 5. COMPUTATIONAL DETAILS

For a complete description of Emulsion Stability Simulations the reader is referred to [Urbina-Villalba, 2009; Toro-Mendoza, 2010; Osorio, 2011; Urbina-Villalba, 2015] and references therein. In ESS the displacement of a particle: $\Delta r = \left| \vec{r}_i (t + \Delta t) - \vec{r}_i (t) \right|$ during a time step $\Delta t$ is the product of the potential of interaction between the particles ($F_{ij} = -dV/dr_{ij}$) and the Brownian drift induced by the solvent molecules $\sqrt{2 D_{eff,i} \Delta t}$ [$\vec{G}auss$]:

$$\vec{r}_i (t + \Delta t) = \vec{r}_i (t) + \sum_j (\vec{F}_{ji} \, D_{eff,i} / k_B T) \, \Delta t$$
$$+ \sqrt{2 D_{eff,i} \Delta t} \, [\vec{G}auss] \qquad (22)$$

Here "Gauss" is a three component vector which stands for a set of random variables characterized by a Gaussian distribution, (with zero mean and unit variance), and $D_{eff,i}$ is the effective diffusion constant of particle $i$.

The parameters of the DLVO potential employed in the present simulations (surface charge, Hamaker constant, etc) correspond to the ones previously used in the simulations of sets I (T - I) and II (T - II) of Ref. [Urbina-Villalba, 2015] (Table 1). The resulting potentials corresponding to non-deformable drops are illustrated in Figs. 3 - 5 of the referred article.

While the charge of the drops belonging to T - I vary slightly from 0 to 1 M

Table 1: Surface charge σ (in Coul/m²) of hexadecane in water 184-nm drops stabilized with 0.5 and 7.5 mM SDS according to T - I, T - II or T - III parameterization.

| [NaCl] mM | σ (Coul/m²) T - I, 0.5 mM SDS | σ (Coul/m²) T - II, 0.5 mM SDS | σ (Coul/m²) T - III, 0.5 mM SDS | σ (Coul/m²) T - I, 7.5 mM SDS | σ (Coul/m²) T - II, 7.5 mM SDS | σ (Coul/m²) T - III, 7.5 mM SDS |
|---|---|---|---|---|---|---|
| 0 | -0.02220169 | -0.03392784 | -0.03246798 | -0.03372739 | -0.01363696 | -0.01522633 |
| 100 | -0.03148389 | -0.07434022 | -0.07672295 | -0.03829936 | -0.07323209 | -0.07970597 |
| 200 | -0.03334607 | -0.08523800 | -0.08865695 | -0.03829936 | -0.08466665 | -0.09207772 |
| 300 | -0.03436343 | -0.09180175 | -0.09584482 | -0.03829936 | -0.09141683 | -0.09938116 |
| 325 | -0.03455781 | -0.09310886 | -0.09727621 | -0.03829936 | -0.09275297 | -0.10082681 |
| 350 | -0.03473591 | -0.09432185 | -0.09860454 | -0.03829936 | -0.09399092 | -0.10216623 |
| 400 | -0.03505235 | -0.09651367 | -0.10100477 | -0.03829936 | -0.09622345 | -0.10458173 |
| 450 | -0.03532677 | -0.09845297 | -0.10312846 | -0.03829936 | -0.09819453 | -0.10671437 |
| 475 | -0.03545129 | -0.09934488 | -0.10410518 | -0.03829936 | -0.09909987 | -0.10769391 |
| 500 | -0.03556859 | -0.10019196 | -0.10503280 | -0.03829936 | -0.09995904 | -0.10862350 |
| 550 | -0.03578441 | -0.10176818 | -0.10675890 | -0.03829936 | -0.10155618 | -0.11035155 |
| 600 | -0.03597904 | -0.10320951 | -0.10833727 | -0.03829936 | -0.10301499 | -0.11192992 |
| 650 | -0.03615606 | -0.10453722 | -0.10979123 | -0.03829936 | -0.10435752 | -0.11338249 |
| 700 | -0.03829936 | -0.10576793 | -0.11113897 | -0.03829936 | -0.10560095 | -0.11472783 |
| 800 | -0.03829936 | -0.10798868 | -0.11357087 | -0.03829936 | -0.10784240 | -0.11715299 |
| 900 | -0.03829936 | -0.10995057 | -0.11571931 | -0.03829936 | -0.10982043 | -0.11929315 |
| 1000 | -0.03829936 | -0.11170771 | -0.11764353 | -0.03829936 | -0.11159050 | -0.12120829 |

NaCl (-0.02 to -0.04 Coul/m²), the ones of T - II are substantially higher, and increase progressively from -0.02 to -0.11 Coul/m². There is no substantial difference between the charges predicted using average coefficients (T - II) or concentration dependent coefficients (T - III) for 7.5 mM (see Eq. (29) in Ref. [Urbina-Villalba, 2013], Table 1 and Figure 2).

In the case of deformable droplets the analytical form of the DLVO potential changes with the geometry of the drops. The code transforms spherical drops into truncated spheroids as soon as the initial distance of deformation $h_0$ is reached. The energy of the process implies the inclusion of two additional potentials (extensional and bending) which take into account the increase of the total interfacial area of the drops, and the change in curvature of the surfactant layer adsorbed. Figures 3-6 illustrate the form of the total potential between two deformable droplets for 0.5 and 7.5 mM SDS. The surface excess and surface tension required for the potentials are the ones corresponding to Models 3 (T - I) and 4 (T – II) of Ref. [Urbina-Villalba, 2013]. The value of the bending constant was fixed to $1.6 \times 10^{-12}$ N [Os-orio, 2011]. Spherical drops move with the tensors of Stokes and Honig [Urbina-Villalba, 2013]. Truncated sphere diffuse with the tensor of Danov with $\varepsilon_s = 1$ (see Refs. [Toro-Mendoza, 2010; Osorio, 2011] for details).

In general the potentials of deformable drops have several critical points (Figs. 3-6). The sum of the extensional, bending, and van der Waals potentials resemble a shark´s fin [Toro-Mendoza, 2010]. The potential increases until the maximum film radius is attained. The thinning of the film occurs after the maximum, where the van der Waals interaction predominates (lower separation distances). The combination of the extensional, bending and van der Waals potentials is illustrated in figures 3(b), 4(a), 5(a,b) and 6(a) for a salt concentration of 0 mM NaCl (cyan curves). The DLVO peak is very sharp and can be clearly distinguished on top of the other contributions. In general it is located in the same region for most salt concentrations. At very low salinities, the DLVO barrier is much wider (and higher), generating a secondary minimum that lies substantially farther away than $h_0$.

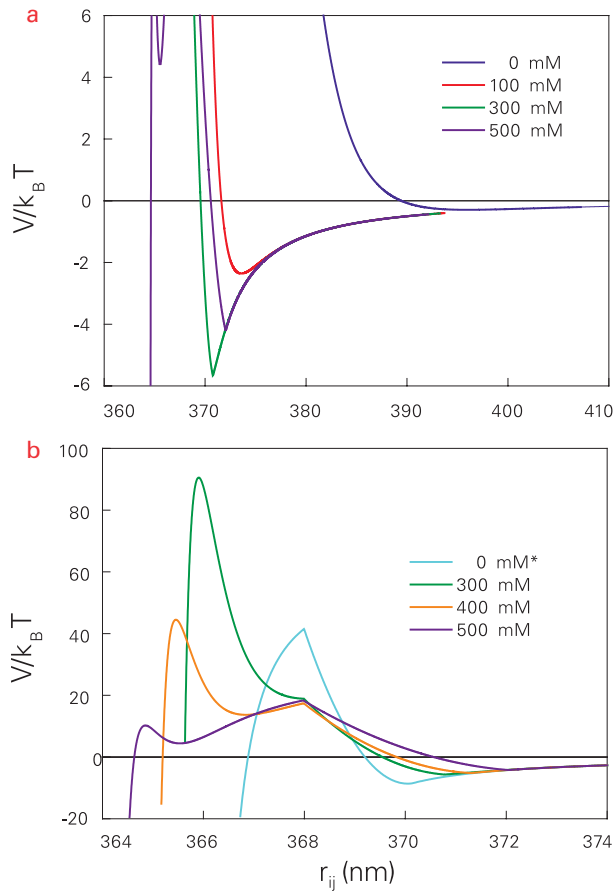

Fig 3: Interaction potential between two deformable 184-nm drops of hexadecane in water. The electrostatic potentials correspond to the predictions of macroscopic adsorption isotherms for a surfactant concentration of 0.5 mM SDS (T - I parameterization). The asterisk corresponds to the total potential in the absence of the electrostatic interaction (van der Waals, extensional and bending potentials).

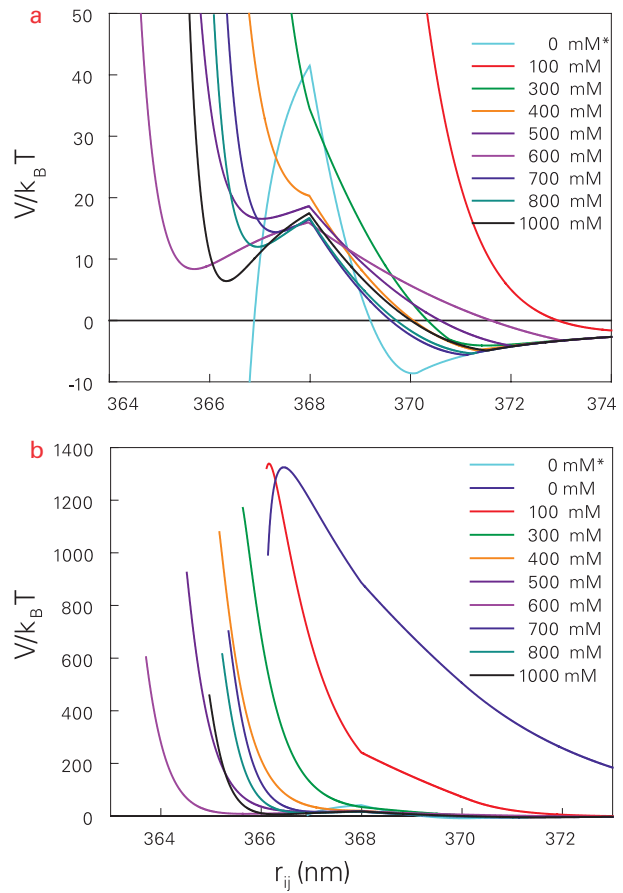

Fig 4: Interaction potential between two deformable 184-nm drops of hexadecane in water. The electrostatic potentials correspond to the predictions of the variation of the zeta potential of the drops as a function of the salt concentration 0.5 mM SDS (T - II). The asterisk corresponds to the total potential in the absence of electrostatic interaction (van der Waals, extensional and bending potentials).

Deformable drops coalesce if they reach a critical separation distance $h_{crit}$. Figures 4(b) and 6(b) illustrate several cases in which the value of $h_{crit}$ is reached before the maximum of the electrostatic interaction is attained (the curves end at $h_{crit}$). However, according to the simulations, coalescence does not occur unless an additional mechanism other than film drainage is implemented.

In a typical DLVO potential, the repulsive barrier lowers and the minimum displaces toward shorter separation as the ionic strength increases. However, if the charge of the particles is generated by surfactant adsorption, a high salt concentration decreases the electrostatic potential but also enhances adsorption increasing the potential and lowering the interfacial tension. Consequently, the position of the sec-ondary minimum (and its depth) in the case of deformable droplets does not change in a monotonous way due to the particular behavior of the extensional and bending potentials as a function of the salt concentration (see Fig. 4(a)). Moreover, the confluence of the electrostatic potential and the van der Waals interaction at very short separations generates and additional (tertiary) minimum. There are cases in which de electrostatic barrier develops at even shorter separation distances (Fig. 4(a, b)). In these cases the barrier tends to vary as predicted by Debye-Hückel theory.

In general, the dependence of the potential on the surface excess alters the order expected by the DLVO theory in terms of the ionic strength (see the curves corresponding to 700-1000 mM in Fig. 4(b)). In any event, the tertiary mini-mum was never reached in these simulations. Thus, for the present purposes, only flocculation with respect to the outermost (secondary) minimum is considered.

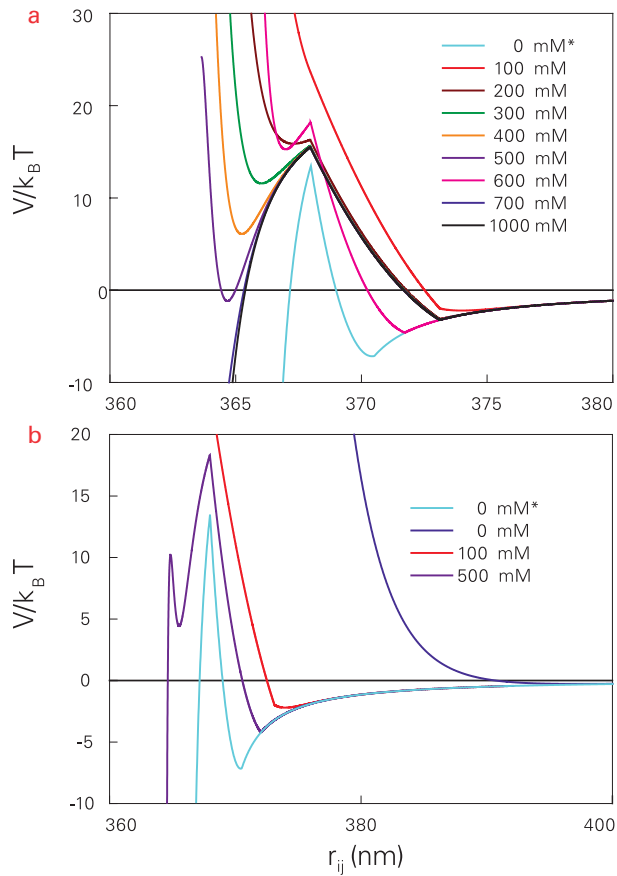

Fig 5: Interaction potential between two deformable 184-nm drops of hexadecane in water. The electrostatic potentials correspond to the predictions of microscopic adsorption isotherms for a surfactant concentration of 7.5 mM SDS (T - I parameterization).

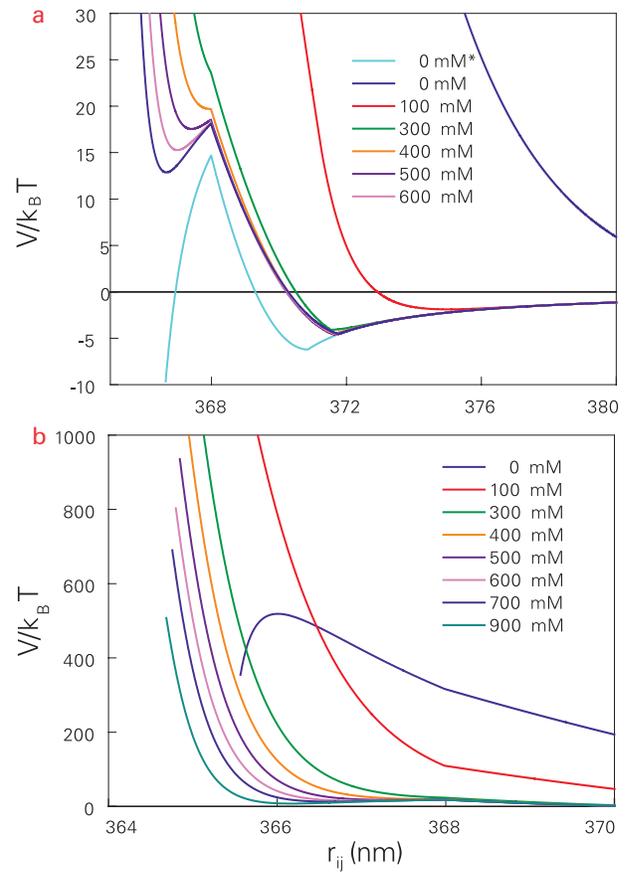

Fig 6: Interaction potential between two deformable 184-nm drops of hexadecane in water. The electrostatic potentials correspond to the predictions of the variation of the zeta potential of the drops as a function of salt, for a surfactant concentration of 7.5 mM SDS (T - II parameterization).

For two-particle simulations a box of $L = 5$ R was used. The drops were located along the x-axis. The initial separation distance was either 4 nm (spherical drops) or 8 nm (deformable drops) except for 0 - 100 mM NaCl (see Table 2).

A long preliminary study was necessary in order to establish an adequate criterion for secondary minimum flocculation. For this purpose the following variables were introduced as input: a) a flocculation distance ($d_{floc}$), b) a sampling time ($t_s$), c) the percentage of time spent at a distance lower than $d_{floc}$, and d) a minimum number of iterations prior to the calculation of the referred percentage. It became evident that in order to measure a relevant time of residence in the well of the potential, the separation dis-tance between the drops should be referenced to the location of the secondary minimum; $d_{floc} = d_{min}$ (Table 2). The time ($\tau_f$) spent by the particles at a distance of separation ($h = r_{ij} - R_i - R_j$) shorter than the position of the secondary minimum can be computed, adding the number of time steps ($\Delta t$) for which $h \leq d_{floc}$:

$$t_f = \sum_{i=1}^{i=N_m} \Delta t \, \delta_i (h \leq d_{floc}) \tag{23}$$

Where the $\delta_i$ function is equal to 1 if $h \leq d_{floc}$ in step $i$, or 0 otherwise. After every iteration, $t_f$ is compared with an arbitrary preselected sampling time ($t_s$) introduced as an input of the simulation. When $t_f = t_s$ the program writes the total computation time elapsed ($t_{tot,m}$):

$$t_{tot,m} = \sum_{t=1}^{i=N_m} \Delta t = N_m \Delta t \tag{24}$$

Table 2: Minimum flocculation distance $d_{floc}$ (nm), used in the simulation of hexadecane in water emulsions stabilized with 0.5 and 7.5 mM SDS.

| [NaCl] mM | $d_{floc}$ (nm) 0.5 mM T - I | $d_{floc}$ (nm) 0.5 mM T- II | $d_{floc}$ (nm) 7.5 mM T - I | $d_{floc}$ (nm) 7.5 mM T - II |
|---|---|---|---|---|
| 0 | 29.48 | 29.26 | 25.75 | 26.30 |
| 100 | 7.01 | 7.05 | 6.99 | 7.09 |
| 200 | 4.59 | 4.62 | 4.58 | 4.66 |
| 300 | 3.51 | 3.55 | 3.51 | 3.59 |
| 325 | 3.32 | 3.37 | 3.32 | 3.39 |
| 350 | 3.16 | 3.20 | 3.16 | 3.23 |
| 400 | 2.88 | 2.92 | 2.87 | 2.94 |
| 450 | 2.64 | 2.68 | 2.64 | 2.71 |
| 475 | 2.54 | 2.59 | 2.54 | 2.61 |
| 500 | 2.45 | 2.48 | 2.45 | 2.51 |
| 550 | 2.28 | 2.32 | 2.28 | 2.35 |
| 600 | 2.13 | 2.17 | 2.13 | 2.20 |
| 650 | 2.01 | 2.04 | 2.01 | 2.07 |
| 700 | 1.90 | 1.93 | 1.90 | 1.96 |
| 800 | 1.70 | 1.74 | 1.70 | 1.77 |
| 900 | 1.54 | 1.58 | 1.54 | 1.60 |
| 1000 | 1.41 | 1.44 | 1.41 | 1.47 |

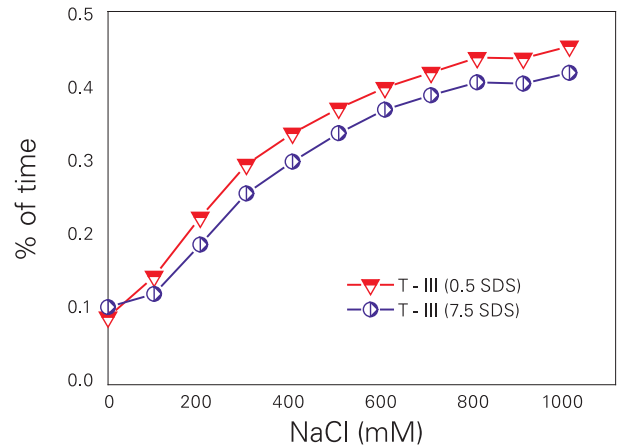

Fig. 7: Percentage of time spent by two flocculated drops at a distance shorter than $d_{floc}$.

$N_m$ is the number of cycles of iteration of the run "$m$" needed to reach $t_f = t_s$ (subscript "m" is an integer that identifies each individual simulation). Several values of $t_s$ were formerly proven, but only two were selected (0.01, and 1 s) in order to check the insensibility of the results with respect to the magnitude of this parameter.

The procedure outlined above was repeated a given number of times ($N_{max} = 100$) using different seeds for the random number generator in order to promote different trajectories:

$$ t_{ave} = \frac{1}{N_{max}} \sum_{m=1}^{N_{max}} t_{tot,\, m} \qquad (25) $$

The average time spent by the particles below $d_{floc}$ depends on the curvature of the potential near a repulsive barrier. The "hardness" of the potential (slope at h < $d_{min}$) increases abruptly with the height of the barrier. In the case of non-deformable drops, the barrier decreases and the depth of the secondary minimum increases with the increase of the ionic strength. Thus, the time spent by the particles around $d_{floc}$ increases considerably, and the sampling time ($t_f = t_s$) is reached sooner. Figure 7 shows the percentage of time spent by two spherical particles below $d_{floc}$ after 2 billion iterations. The percentage increases steadily until 800 mM NaCl for both 0.5 and 7.5 mM SDS.

Using Eq. (21):

$$ k_f^{slow} = \left( \frac{t_{ave}^{fast}}{t_{ave}^{slow}} \right) k_f^{fast} \qquad (26) $$

Thus, the average times in the parenthesis can be evaluated using two-particle simulations. The "fast" terms are computed neglecting the electrostatic repulsion of the DLVO potential, but keeping the extensional and bending potentials in the case of deformable drops.

The same criteria used for two-particle calculations (Eqs. (23)-(25)) was employed in the corresponding many-particle simulation needed to appraise $k_{FC}^{fast}$. In this case the particles were artificially coalesced when $t_f = t_s$. Hence, the number of particles artificially decreases with time as a result of secondary minimum flocculation.

The purpose of this article is to test the influence of drop deformation in the flocculation rate of hexadecane-in-water emulsions using two-particle calculations (Eq. (26)). Several types of simulations (S) were run:

a) SD: Non-deformable Spherical Drops. These runs are meant to reproduce the qualitative behavior of previous many-particle simulations for 0.5 mM SDS (Figs. 1) in order to test the soundness of the two-particle method-ology. In order to avoid unnecessary computations, the values of $k_{FC}^{fast}$ previously obtained for a salt concentration of 1000 mM were used: $k_{FC}^{fast} = 2.54 \times 10^{-18}$ m$^3$/s for 0.5 SDS and 2.86 x 10$^{-18}$ m$^3$/s for 7.5 mM SDS [Urbina-Villalba, 2015].

b) DD1: Deformable Drops. Both T - I and T - II parameterizations were tested. This means that the surface charge was calculated either by using adsorption isotherms or the zeta potential of the drops. The tensor used to move the drops resemble the draining of the intervening water film.

c) DD2: Effect of Capillary waves. These calculations assume that coalescence occurs when the height of the surface oscillations is greater than the width of the water film between flocculated drops [Vrij, 1966; Vrij, 1968; Urbina-Villalba, 2009]. The total extent of the oscillations is approximated by:

$$\lambda_T = (\lambda_i + \lambda_j) \exp(\tau_{ij}/\tau_{Vrij}) > h \qquad (27)$$

Where:

$$\lambda_i = Ran(t) * h_{crit} \qquad (28)$$

$$\tau_{Vrij} = 96 \pi^2 \gamma \eta h_0^5 A_H^{-2} \qquad (29)$$

$$\gamma = (\gamma_i + \gamma_j)/2 \qquad (30)$$

In these equations, "Ran" stands for a random real number between -1 and 1, $\gamma$ is the average value of the interfacial tension, and $\tau_{ij}$ is the time of existence of a doublet, which was approximated by $t_f$ (Eq. (23)). Equation (29) was deduced by Vrij considering van der Waals forces only. Analytical expressions including the repulsive potential are not available. Moreover, according to the theory, it is necessary to assure that:

$$\frac{d^2V}{db^2} \leq -\frac{2\pi\gamma}{a^2} \qquad (31)$$

in the presence of additional repulsions. In Eq. (31) $a$ represents the dimension (~ radius) of the film. This condition typically occurs near the primary minimum, and also at distances larger than the position of the secondary minimum (see Fig. 3 in Ref. [Vrij, 1966]). The algorithm implemented in the program (Eqs. (27) – (30)), does not consider the restriction imposed by Eq. (31).

d) DD3: Hole formation [de Vries, 1958]. In the case of a foam, the activation energy needed to expand a hole in the liquid film is equal to:

$$E_{de Vries} = 0.73 h^2 \gamma \qquad (32)$$

Here, Eq. (32) is used to mimic the appearance of a hole across the water film between aggregated drops. The value of $E_{de Vries}$ is compared with the sum of the kinetic energy provided by the solvent to each drop whenever $h < h_0$. Notice, that since the algorithm is based on Brownian Dynamics, it resembles the movement of the particles in the diffusion regime (Eq. (22)). Therefore the instantaneous velocity of a drop produced by the momentum exchange with the solvent molecules is approximated by: $\left| \sqrt{6D_{eff,i} \Delta t} \left[ \vec{Gauss} \right] \right| / \Delta t$. Hence, the fluctuations of the velocity depend (artificially) on the time step of the simulation. A small time step of 7.94 x 10$^{-10}$ s was *arbitrarily* selected in order to appreciate the qualitative effect of large fluctuations in the process of hole formation. This corresponds to an average velocity of 32.7 m/s ($54 k_B T$) in 5 million iterations. In order to reproduce the thermal velocity of a 184-nm drop in the ballistic regime (~ 11.5 mm/s) a time step of 7.14 x 10$^{-9}$ s is required. Notice that the time step used in the rest of the simulations ($\Delta t = 7.94$ x $10^{-8}$ s), as well as in previous many-particle calculations [Urbina-Villalba, 2015], **corresponds to the diffusion regime**.

## 6. RESULTS AND DISCUSSION

Figs. 8 ilustrates the outcome of SD simulations for 0.5 and 7.5 mM SDS. The results for concentration-independent coefficients and the ones corresponding to 7.5 mM are similar. This is not surprising considering the small differences in the surface charge of the drops (Table 1) suggested by Eq. (29) of Ref. [Urbina-Villalba, 2013]. SD calculations show concave downwards curves whose slope progressively diminishes. This behavior is fairly similar to the performance of equivalent many-particle simulations (Figure 2). Consequently, the two-particle calculations reproduce the qualitative of the experimental data corresponding to 0.5 mM SDS (Fig. 1), but differ considerably from the data belonging to 7.5 mM SDS (Fig. 2). The rates of SD calculations increase monotonously as a function of the salt concentration. Thus, SD simulations suggest that only for 0.5 mM SDS, hexadecane drops of 184 nm behave as spherical non-deformable drops. However, SD calculations also suggest faster aggregation rates (the theo-retical curves systematically lie $(0.5 - 1.0) \times 10^{18}$ m$^3$/s units above the experimental data). Moreover, the slope of the curves is similar but not parallel to the one of the experimental plot. It is remarkable that all curves (theoretical and experimental) show the same inflexion point at 900 mM SDS.

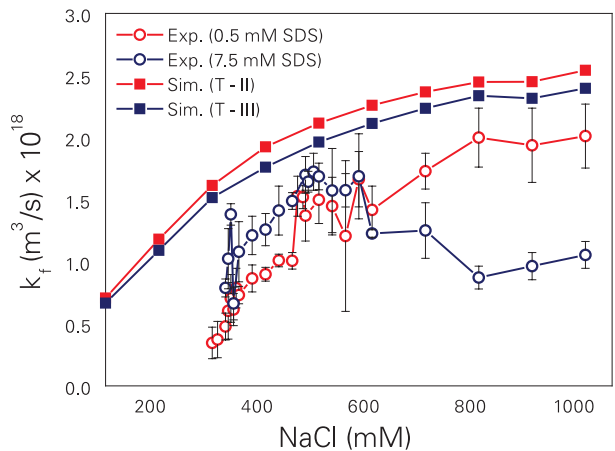

Fig. 8: Flocculation rates predicted by SD simulations (spherical non-deformable drops) with T - II or T - III parameterization corresponding to a set of hexadecane in water nanoemulsions stabilized with either 0.5 or 7.5 mM SDS.

If deformable drops are used to simulate the behavior of the 0.5 mM-systems (DD1 simulations) drastic discrepancies are observed. The predictions differ quantitatively and qualitatively from the experimental data. The theoretical curves show two maxima and a minimum value around 600 mM NaCl. The rates resulting from T - I parameterization appear to be independent of the salt concentration above 600 mM while the ones of T - II increase up to 700 mM SDS and decrease afterwards. This behavior closely follows the position and depth of the corresponding secondary minima. The rate of flocculation is slower when the secondary minimum is shallower and occurs at a greater distance of separation between the particles. It also decreases with the steepness of the repulsive barrier close to the minimum. Thus, the results of DD1 simulations for **0.5 SDS** do not follow the experimental trend, suggesting that the drops of these systems **are not deformable**. This fact confirms the previous findings of SD calculations.

Unlike the previous case, the simulations of spherical drops (SD) corresponding to **7.5 mM SDS** (high surfactant concentration) fail to reproduce the experimental data (Fig. 8). Two additional factors may influence this behavior. First, the **surfactant concentration is higher than the critical micelle concentration (CMC) for all** salinities studied (Table 3). Second, the **drops are more deformable** due to the lower value of the interfacial tension.

The effect of micelles cannot be appraised with this type of calculations [Ariyaprakai, 2007], but it is likely that micelles increase the solubilization of the oil with the augment of the ionic strength, since the CMC decreases with the amount of salt. Whether this process occurs or not during the 50 seconds of the experiments is uncertain. If it would, the average radius of the emulsions should decrease with the ionic strength inducing slower aggregation rates.

The results of DD1 simulations markedly depend on the method of parameterization for a surfactant concentration of 7.5 mM SDS. Systems parameterized with adsorption isotherms (T - I) suggest a value of $k_f$ which is fairly insensible to a change of the ionic strength. The theoretical points cross the experimental curve horizontally intercepting it at four different salt concentrations. Instead, T - II calculations show a smooth concave curve with an ample maximum (between 500 and 600 mM NaCl). This behavior reflects similar secondary minima. While T - I parametrization predicts the saturation of the interface for all salinities higher than 100 mM, T - II parametrization favors a smooth variation. The minima occurring at 300 mM for DD1 (T - I) simulations reflect that the corresponding secondary minimum is less deep but occurs at a shorter distance of separation than the rest, of the systems, favoring a considerably slower flocculation rate.

Both the DD1 (T - II) and the DD2 (T - II) simulations follow the qualitative behavior of the experimental data, but exhibit substantially higher aggregation rates. This might be an indication of an inaccurate estimation of $k_{FC}^{fast}$. Note that the use of Eq. (29) in DD2 simulations induces faster coalescence times. Besides, the time of residence in the secondary minimum evaluated by means of Eq. (23) favors shorter aggregation times (larger rates) since the effect of the repulsive barrier is partially disregarded. This (possible) limitation can be corrected requiring the continuous permanence of the particles in the minimum for the calculation of $t_f$. These calculations are on their way, but they require substantially longer computation times. Contrary to our expectations, the many-particle calculations run substantially faster in this case than the two-particle simulations. This might indicate that a more involved implementation of the random number simulator is required: Every time a particle enters the secondary minimum it is quickly repelled. In order to accumulate a substantial time of residence within the minimum, the random component must be very small, or the provided impulse should favor the approach of the particles, nullifying the effect of the potential. In a 500-particle calculation, the generator must produce 499 x 3 numbers before going back to the same particle. Instead, in a two-particle simulation it goes back to the same particle after 3 iterations. According to these preliminary results, it appears that it is more probable to produce an "adequate" random deviate for flocculation in the first case. Consequently, the technique of two-particle simulations does not reduce the computation time in this particular case.

| [NaCl] mM | SDS CMC (mM) |
|---|---|
| 100 | 1.62 |
| 200 | 1.18 |
| 300 | 0.98 |
| 325 | 0.94 |
| 350 | 0.91 |
| 400 | 0.86 |
| 450 | 0.81 |
| 475 | 0.79 |
| 500 | 0.77 |
| 550 | 0.74 |
| 600 | 0.71 |
| 650 | 0.69 |
| 700 | 0.66 |
| 800 | 0.62 |
| 900 | 0.59 |
| 1000 | 0.56 |

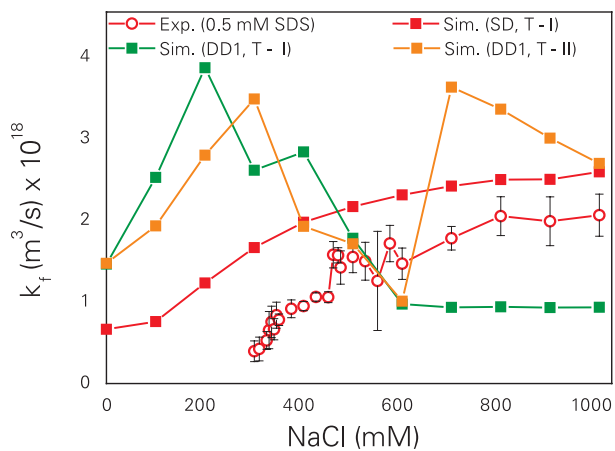

**Fig. 9:** Flocculation rates predicted by DD1 simulations (deformable drops) with T - I and T - II parameterization for a set of hexadecane in water nanoemulsions stabilized with 0.5 mM SDS.

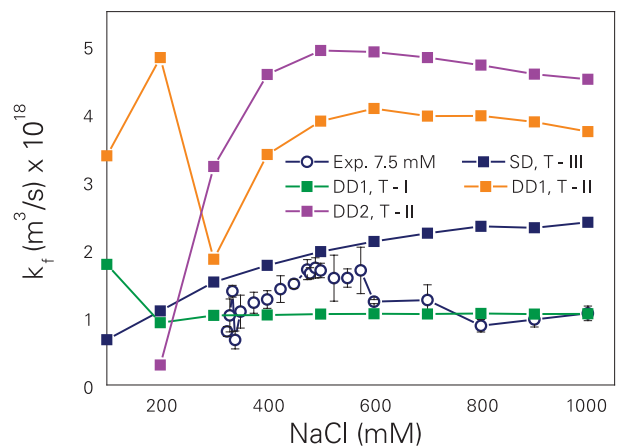

**Fig. 10:** Flocculation rates predicted by DD1, and DD2 simulations with T - I and T - II parameterization for a set of hexadecane in water nanoemulsions stabilized with 7.5 mM SDS.

Finally, DD3 simulations (de Vries) also suggest a fairly constant aggregation rate as in the case of DD1 (T - II) sim-ulations, but in this case, an average rate of 1.5 x $10^{-17}$ m$^3$/s results (not shown). This is partially the consequence of the small time step employed which was purposely selected to emphasize the effect of the fluctuations in the diffusive regime. Unfortunately, it apparently induced unexpectedly short coalescence times. Longer time steps will reduce the magnitude of the thermal fluctuations favoring slower aggregation rates (see Section 5). Hence, additional simulations are required. As in the case of the surface oscillations, longer time steps will substantially increase the computation time of the two-particle calculation, limiting their usefulness as a prediction tool.

## 7. CONCLUSIONS

According to these results, the agreement between two-particle simulations and the experimental data is only qualitative but of the correct order of magnitude. The present calculations suggest that the drops of hexadecane in water behave as non-deformable particles at a surfactant concentration of 0.5 mM, but they are probably deformable in an aqueous solution of 7.5 mM SDS. However, while the calculations overwhelming discard the occurrence of deformable drops at 0.5 mM SDS, the results of the 7.5 mM simulations are less persuasive. From the quantitative point of view, the calculation that resembles the experimental behavior more closely (DD1) is the one parameterized with adsorption isotherms (T - I). This contradicts the results of many-particle and two-particle simulations for 0.5 mM SDS, which favors a T - II parameterization. DD1 (T - I) calculations suggest the independence of the flocculation rate with respect to the ionic strength above 100 mM NaCl. From the qualitative point of view, the simulations parameterized with zeta potential measurements (DD1 and DD2) show a broad maximum which occurs approximately at the same concentration of the small maximum experimentally found, but in the latter case the theoretical rates are substantially faster.

In order to establish a definite conclusion regarding coalescence and deformability, it appears necessary to: a) confirm the experimental variation of the 7.5 mM SDS systems with an independent set of measurements; b) account for the effect of micelles; c) evaluate the effect of a continuous time of residence for secondary minimum flocculation; and d) implement more accurate ways for calculating $k_{FC}^{fast}$.

# 6. ACKNOWLEDGEMENT

The technical assistance of Dr. Kareem Rahn-Chique in the elaboration of the figures and the correction of the manuscript is gratefully acknowledged.